\documentclass{PoS}

\usepackage{amsmath}
\usepackage{amssymb}
\usepackage{color}
\usepackage{graphicx}
\usepackage{multirow}
\usepackage{float}

\title{Lattice study of the Schwinger model at fixed topology}

\ShortTitle{Lattice study of the Schwinger model at fixed topology}

\author{\speaker{Christopher Czaban} \\
       Goethe-Universit\"at Frankfurt am Main, Institut f\"ur Theoretische Physik, \\ Max-von-Laue-Stra{\ss}e 1, D-60438 Frankfurt am Main, Germany \\
       E-mail: \email{czaban@th.physik.uni-frankfurt.de}}
       
\author{Marc Wagner \\
       Goethe-Universit\"at Frankfurt am Main, Institut f\"ur Theoretische Physik, \\ Max-von-Laue-Stra{\ss}e 1, D-60438 Frankfurt am Main, Germany \\
       E-mail: \email{wagner@th.physik.uni-frankfurt.de}}

\abstract{At small lattice spacing QCD simulations are expected to become stuck in a single topological sector. Observables evaluated in a fixed topological sector differ from their counterparts in full QCD, i.e. at unfixed topology, by volume dependent corrections. We investigate these corrections in the two-flavor Schwinger model, which is in several aspects similar to QCD, using Wilson fermions. We also try to remove these corrections by suitable extrapolations to infinite volume.}

\FullConference{31st International Symposium on Lattice Field Theory LATTICE 2013 \\
		 July 29--August 3, 2013 \\
		 Mainz, Germany}

\newcommand{\ltapprox}{\raisebox{-0.5ex}{$\,\stackrel{<}{\scriptstyle\sim}\,$}}

\begin{document}


\section{Introduction}

Lattice QCD simulations are expected to suffer from frozen topology at small values of the lattice spacing independent of the quark discretization. The reason is that gauge link configurations belong to different topological sectors, which are separated by barriers of rather large Euclidean action. Choosing a smaller value for the lattice spacing increases these barriers, until standard HMC simulation algorithms are not anymore able to frequently traverse these barriers. Then the simulation gets stuck in a certain topological sector for a rather long time and computed observables contain corresponding systematic errors (cf.\ e.g.\ \cite{Luscher:2011kk} and references therein). When using overlap fermions, topology freezing is even observed at rather coarse lattice spacings \cite{Aoki:2008tq}.

A possible solution to these problems is to restrict computations to a single topological sector, either by sorting the generated gauge link configurations with respect to their topological charge or by directly employing so-called topology fixing actions (cf.\ e.g.\ \cite{Fukaya:2005cw,Bietenholz:2005rd,Bruckmann:2009cv}). In a second step systematic effects due to topology fixing need to be removed by suitable extrapolations. Corresponding expressions have been derived \cite{Brower:2003yx,Aoki:2007ka} and tested in simple models, i.e.\ in the Schwinger model \cite{Bietenholz:2012sh} and in quantum mechanics \cite{Dromard:2013wja}.

In this work we explore computations at fixed topology in the Schwinger model. In contrast to \cite{Bietenholz:2012sh}, we use a computationally much cheaper lattice discretization of fermions (Wilson fermions instead of overlap fermions), which allows, to generate lattice results for many different topological sectors and spacetime volumes. In addition to the pseudoscalar meson mass (the ``pion mass'') we also study the static potential.


\section{The Schwinger model with $N_f=2$ flavors of fermions}


\subsection{The Schwinger model in the continuum}

The Schwinger model describes 2-dimensional Euclidean quantum electrodynamics:
\begin{eqnarray}
\label{EQN001} \mathcal L(\psi,\bar\psi,A) \ \ = \ \ \sum_{f=1}^{N_f} \bar\psi^{(f)} \Big(\gamma_\mu (\partial_\mu + i g A_\mu) + m\Big) \psi^{(f)} + \frac{1}{4} F_{\mu\nu} F_{\mu\nu} .
\end{eqnarray}
It is a well known toy model for QCD, since it shares several interesting features with QCD. For example the $U(1)$ gauge theory in two spacetime dimensions allows for topologically non-trivial field configurations, which are similar to instantons in $4$-dimensional Yang-Mills theory. The corresponding topological charge is given by
\begin{eqnarray}
Q \ \ = \ \ \frac{1}{\pi} \int \text{d}^2x \, \epsilon_{\mu\nu} F_{\mu\nu} .
\end{eqnarray}
Moreover, for $N_f = 2$ its low lying energy eigenstates contain a rather light iso-triplet, which are quasi Nambu-Goldstone bosons and, therefore, can be seen as the pions of this model. Finally, the model provides fermion confinement.


\subsection{The Schwinger model on the lattice}

We study the Schwinger model (\ref{EQN001}) on a periodic spacetime lattice with $N_L^2$ lattice sites corresponding to a spacetime extension of $L = N_L a$ ($a$ is the lattice spacing) and a spacetime volume $V = L^2$. We use $N_f = 2$ flavors of Wilson fermions and the Wilson plaquette gauge action. As usual all dimensionful quantities are expressed in units of $a$ and denoted by $\hat{\phantom{x}}$, e.g.\ $\hat{g} = g a$ and $\hat{m} = m a$.

One can approach the continuum limit by increasing $N_L$, while keeping the dimensionless ratios $g L = \hat g N_L$ and $M_\pi L = \hat M_\pi N_L$ fixed ($M_\pi$ denotes the mass of the aforementioned quasi Nambu-Goldstone bosons, i.e.\ the pion mass). This requires to decrease both $\hat g$ and $\hat M_\pi$ proportional to $1 / N_L$ (for the latter $\hat{m}$ has to be adjusted appropriately). It is also common to use the dimensionless inverse squared coupling constant $\beta = 1/\hat g^2$.

We use the geometric definition of topological charge on the lattice,
\begin{eqnarray}
\label{EQN432} Q \ \ = \ \ \frac{1}{2 \pi} \sum_P \phi(P)
\end{eqnarray}
\cite{Luscher:1981zq,Gattringer:1997qc}, where $\sum_P$ denotes the sum over all plaquettes $P = e^{i \phi(P)}$ with $-\pi < \phi(P) \leq +\pi$. With this definition $Q \in \mathbb{Z}$ for any given gauge link configuration.

\begin{figure}[H]
\begin{center}
\vspace{-2cm}
\rotatebox{0}{\includegraphics[width=0.6\linewidth]{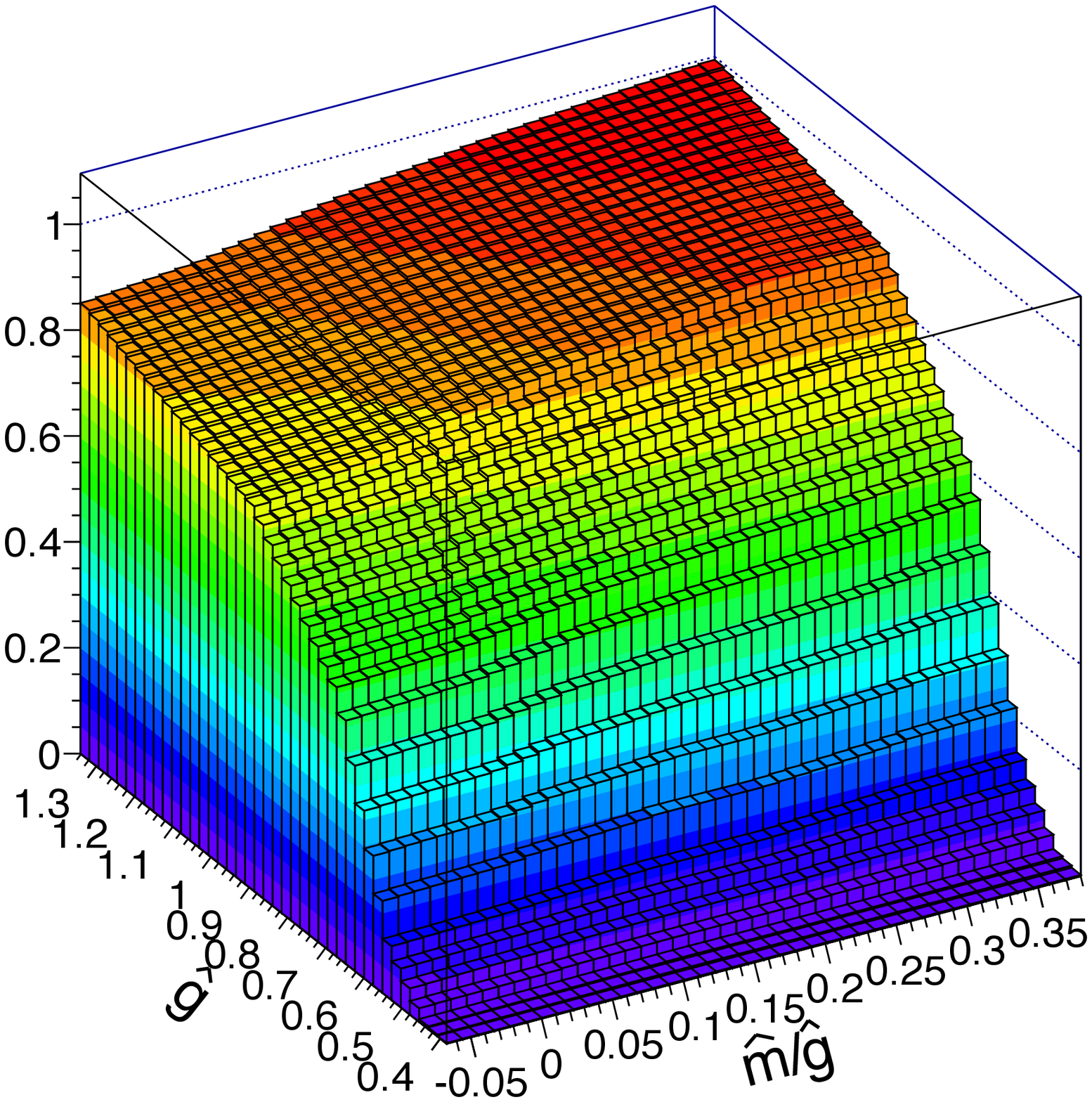}}
\vspace{-2cm}
\caption{\label{FIG001}the probability for a transition to another topological sector per HMC trajectory as a function of $\hat{g} = 1 / \sqrt{\beta}$ and $\hat{m} / \hat{g} = \hat{m} \sqrt{\beta}$.}
\end{center}
\end{figure}

We performed simulations at various values of $\beta$, $\hat{m}$ and $N_L$ using a Hybrid Monte Carlo (HMC) algorithm with multiple timescale integration and mass preconditioning \cite{HMC_Urbach}. In Figure~\ref{FIG001} the probability for a transition to another topological sector per HMC trajectory is plotted versus $\hat{g} = 1 / \sqrt{\beta}$ and $\hat{m} / \hat{g} = \hat{m} \sqrt{\beta}$, while $g L = \hat{g} N_L = N_L / \sqrt{\beta} = 24 / \sqrt{5}$ is kept constant. $\hat{g}$ is proportional to the lattice spacing $a$. $\hat{m} / \hat{g}$ is proportional to $\hat{m} / a$ and, therefore, proportional to $m$, the bare quark mass in physical units. As expected there are frequent changes of the topological sector at large values of the lattice spacing $a$ (large values of $\hat{g}$), while at small values of $a$ (small values of $\hat{g}$) topology freezing is observed. The dependence of the probability for a transition on the bare quark mass $\hat{m} / \hat{g}$ is rather weak.

\section{Physical hadron masses from computations at fixed topology}


\subsection{Basic principle}

The temporal correlation function of a hadron creation operator $O$ at fixed topological charge $Q$ and finite spacetime volume $V$ is
\begin{eqnarray}
\nonumber & & \hspace{-0.7cm} C_{Q,V}(t) \ \ = \ \ \frac{1}{Z_{Q,V}} \int D\psi \, D\bar{\psi} \, \int DA \, \delta_{Q,Q[A]} O^\dagger(t) O(0) e^{-S[\psi,\bar{\psi},A]} \quad , \\
 & & \hspace{0.675cm} Z_{Q,V} \ \ = \ \ \int D\psi \, D\bar{\psi} \, \int DA \, \delta_{Q,Q[A]} e^{-S_E[\psi,\bar{\psi},A]} .
\end{eqnarray}
For sufficiently large $V$ one can use a saddle point approximation and expand the correlation function according to
\begin{eqnarray}
\label{eq:CQ_} & & \hspace{-0.7cm} C_{Q,V}(t) \ \ = \ \ A_{Q,V} e^{-M_{Q,V} t} \\
\label{eq:MQ0} & & \hspace{-0.7cm} M_{Q,V} \ \ = \ \ M(0) + \frac{M''(0)}{2 \chi_t V} \bigg(1 - \frac{Q^2}{\chi_t V}\bigg) + \mathcal{O}(1/V^2)
\end{eqnarray}
\cite{Brower:2003yx,Aoki:2007ka}, where the expansion is in the three parameters $M''(0) t / \chi_t V$, $1 / \chi_t V$ and $Q^2 / \chi_t V$. $M_{Q,V}$ is the mass of the hadron excited by $O$ at fixed topological charge $Q$ and finite spacetime volume $V$, $M(\theta)$ is the hadron mass in a $\theta$-vacuum at infinite $V$ (cf.\ e.g.\ \cite{Coleman:1978ae}), $M(0) = M(\theta = 0)$ is the physical hadron mass (i.e.\ the hadron mass at unfixed topology), $M''(0) = d^2M(\theta) / d\theta^2|_{\theta = 0}$ and $\chi_t$ denotes the topological susceptibility.

To determine physical hadron masses (i.e.\ hadron masses at unfixed topology) from fixed topology computations we use a method, which has been proposed in \cite{Brower:2003yx} and tested in \cite{Bietenholz:2012sh,Dromard:2013wja}:
\begin{enumerate}
\item Perform simulations at fixed topology for different topological charges $Q$ and spacetime volumes $V$, for which the expansion (\ref{eq:CQ_}) and (\ref{eq:MQ0}) is a good approximation, i.e.\ where $M''(0) t / \chi_t V$, $1 / \chi_t V$ and $Q^2 / \chi_t V$ are sufficiently small. Determine $M_{Q,V}$ using (\ref{eq:CQ_}) for each simulation.

\item Determine the physical hadron mass $M(0)$ (the hadron mass at unfixed topology and infinite spacetime volume), $M''(0)$ and $\chi_t$ by fitting (\ref{eq:MQ0}) to the masses $M_{Q,V}$ obtained in step~1.
\end{enumerate}


\subsection{Numerical results}

The hadron masses we investigate in the following are the pion mass $M_\pi$ and the static potential $\mathcal{V}_{\bar{q} q}(r)$ (the ground state energy of a static quark antiquark pair at separation $r$). Suitable hadron creation operators are
\begin{eqnarray}
O_\pi \ \ = \ \ \sum_x \bar{\psi}^{(u)}(x) \gamma_1 \psi^{(d)}(x)
\end{eqnarray}
($\sum_x$ denotes a sum over space and $u$ and $d$ label the two degenerate fermion flavors) and
\begin{eqnarray}
O_{\bar{q} q} \ \ = \ \ \bar{q}(x_1) U(x_1,x_2) q(x_2) \quad , \quad r \ \ = \ \ |x_1 - x_2| ,
\end{eqnarray}
($\bar{q}$ and $q$ represent scalar static color charges and $U(x_1,x_2)$ is the product of spatial links connecting $x_1$ and $x_2$).

We obtain hadron masses $\hat{M}_{Q,V} \equiv \hat{M}_{\pi,Q,V}$ and $\hat{M}_{Q,V} \equiv \hat{\mathcal{V}}_{\bar{q} q,Q,V}(r)$ at fixed topology by first determining the topological charge $Q$ on each gauge link configuration according to (\ref{EQN432}). Then we perform independent computations of the pion mass and the static potential using only gauge link configuration with the same absolute value of $Q$.

\begin{figure}[htb]
\begin{center}
\vspace{-2.0cm}
\rotatebox{90}{\includegraphics[width=0.75\linewidth]{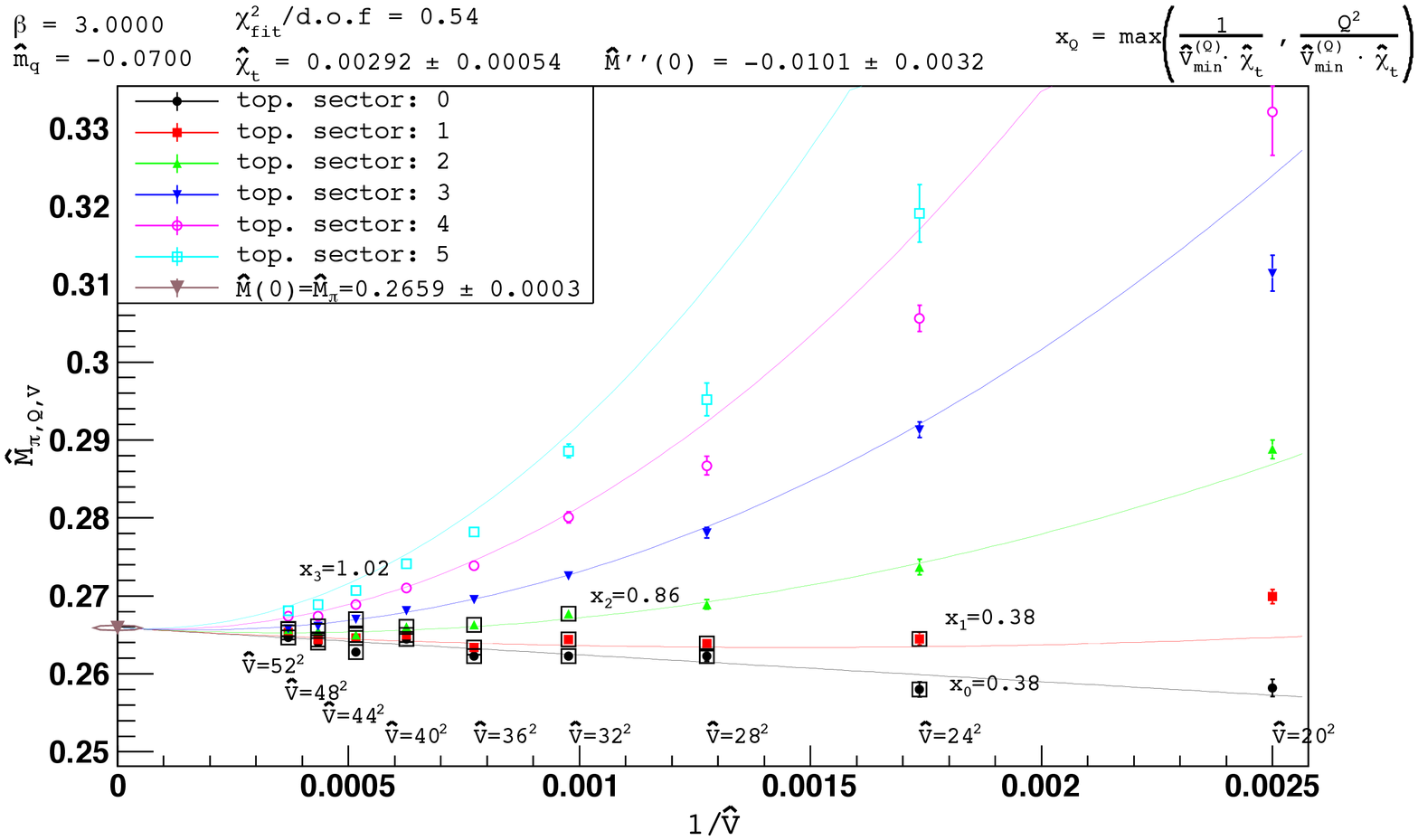}}
\caption{\label{FIG002}lattice results $\hat{M}_{Q,V} \equiv \hat{M}_{\pi,Q,V}$ plotted against $1 / \hat{V}$ for $\beta = 3.0$ and the $\chi^2$ minimizing fit with (3.3); the lattice results $\hat{M}_{Q,V}$ entering the fit are required to fulfill $1 / \chi_t V , Q^2 / \chi_t V \stackrel{<}{\scriptstyle\sim} 1.0$.}
\end{center}
\end{figure}

In Figure~\ref{FIG002} lattice results for $\hat{M}_{Q,V} \equiv \hat{M}_{\pi,Q,V}$ for $\beta = 3.0$, $N_L = 20, \ldots ,52$ and $Q = 0, 1,  \ldots , 5$ are plotted against $1 / \hat{V} = 1 / N_L^2$. These results have been used as input for a single $\chi^2$ minimizing fit with (\ref{eq:MQ0}). As discussed above (\ref{eq:MQ0}) is an expansion in $M''(0) t / \chi_t V$, $1 / \chi_t V$ and $Q^2 / \chi_t V$. Therefore, only fixed topology pion masses $\hat{M}_{Q,V} \equiv \hat{M}_{\pi,Q,V}$ with sufficiently small values of $1 / \chi_t V$ and $Q^2 / \chi_t V$ should be included in the fit. In Figure~\ref{FIG002} we require $1 / \chi_t V , Q^2 / \chi_t V \ltapprox 1.0$ (as indicated by the black boxes)\footnote{The topological susceptibility can be obtained numerically according to $\hat{\chi}_t = \langle Q^2 \rangle / \hat{V}$.}. The fit is of acceptable quality and yields consistent results, $\chi^2/\textrm{dof} = 0.54$, $\hat{M}(0) = \hat{M}_\pi = 0.2659(3)$, which indicates that $1 / \chi_t V , Q^2 / \chi_t V \ltapprox 1.0$ is a reasonable constraint. Since $\hat{M}''(0)$ is determined by the fit, one can only check a posteriori, whether also $M''(0) t / \chi_t V$ is small, which is the case ($|M''(0)| t / \chi_t V \ltapprox 0.09$).
For $\beta = 3.0$ there are sufficiently many transitions between different topological sectors, to safely determine $\hat{M}_\pi$ in the conventional way, i.e.\ by computing the corresponding temporal correlation function on all available gauge link configurations (i.e.\ as an average over all topological sectors) at a single sufficiently large $V$. The result, $\hat{M}_\pi = 0.2663(3)$ (at $\hat{V} = 52^2$), agrees with the above results obtained by fixing topology within statistical errors.

We performed similar determinations of $\hat{M}_\pi$ at fixed topology also at other values of $\beta$ and $\hat{m}$, e.g.\ at $\beta = 4.0$ (which corresponds to a lattice spacing smaller by the factor $\sqrt{4/3} \approx 1.15$ compared to $\beta = 3.0$) and $\hat{m} = -0.03$ (a pion masses larger by the factor $\approx 1.19$ compared to $\beta = 3.0$ and $\hat{m} = -0.07$). As before we obtain accurate results, which are in agreement with conventional computations at unfixed topology.

\begin{figure}[htb]
\begin{center}
\vspace{-2.0cm}
\rotatebox{90}{\includegraphics[width=0.75\linewidth]{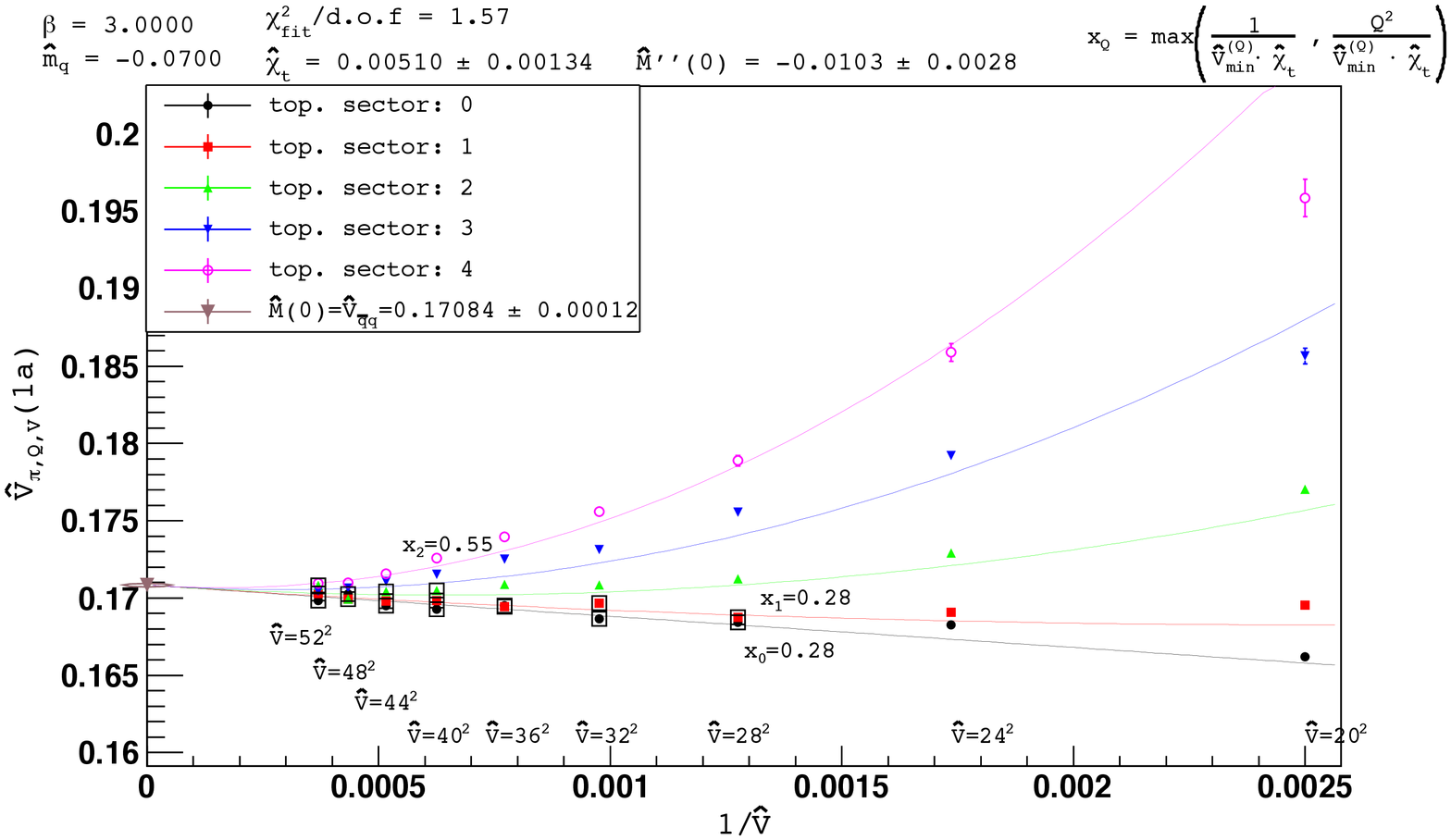}}
\caption{\label{FIG003}lattice results $\hat{M}_{Q,V} \equiv \hat{\mathcal{V}}_{\bar{q} q,Q,V}(a)$ plotted against $1 / \hat{V}$ for $\beta = 3.0$ and the $\chi^2$ minimizing fit with (3.3); the lattice results $\hat{M}_{Q,V}$ entering the fit are required to fulfill $1 / \chi_t V , Q^2 / \chi_t V \stackrel{<}{\scriptstyle\sim} 0.5$.}
\end{center}
\end{figure}

In exactly the same way we successfully determined the static potential at fixed topology. An exemplary plot (analogous to Figure~\ref{FIG002}) corresponding to $\beta = 3.0$, $\hat{m} = -0.07$ and $r = a$ is shown in Figure~\ref{FIG003}.

Further numerical results are collected in Table~\ref{TAB001}. Note that hadron masses (i.e.\ the pion mass and the static potential) can be determined rather precisely (uncertainty $\ll 1 \%$), while there is a rather large error associated with the topological susceptibility (uncertainty up to $\approx 20 \%$). This is in agreement with similar existing investigations in the Schwinger model \cite{Bietenholz:2012sh} and in quantum mechanics \cite{Dromard:2013wja}.

\begin{table}[htb]
\begin{center}
\begin{tabular}{|c|c|c|c|c|c|c|}
\hline
observable & $\beta$ & $\hat{m}$ & $\hat{M}$ (fixed top.) & $\hat{M}$ (conv.)  & $\hat{\chi}_t$ (fixed top.)    & $\hat{\chi}_t$ ($\langle Q^2 \rangle / \hat{V}$) \\ \hline \hline
$M_\pi$ 		    & \multirow{3}{*}{3.0} & \multirow{3}{*}{-0.07} & 0.2659(3)\phantom{0}     & 0.2663(3)\phantom{0}       & 0.00292(54)        & 	\multirow{3}{*}{0.00454(6)\phantom{0}}     \\ 
$V_{Q\bar Q}(1)$ & 				     & 			                 & 0.1708(1)\phantom{0}     & 0.17108(5) 			 & 0.0051(13)\phantom{0}	  &						      \\  
$V_{Q\bar Q}(2)$ &  				     &  		                 & 0.2914(3)\phantom{0}     &  0.2927(2)\phantom{0}			  & 0.00247(20)    & 					      \\ \hline\hline
$M_\pi$			    & \multirow{3}{*}{4.0} & \multirow{3}{*}{-0.03} & 0.2743(6)\phantom{0}     &  0.2743(3)\phantom{0}			    & 0.00228(39)     &  \multirow{3}{*}{0.00353(14)}       \\ 
$V_{Q\bar Q}(1)$ &  				     & 				         & 0.12552(7)   & 0.12551(4)   			& 0.00313(26)        &					          	\\ 
$V_{Q\bar Q}(2)$ &  				     &  			         & 0.2250(2)\phantom{0}    & 0.2247(2)\phantom{0}    			& 0.00329(15)        & 					                \\ \hline
\end{tabular}
\end{center}
\caption{\label{TAB001}a collection of some numerical results (fixed top.: results obtained by fixed topology computations; conv.: results obtained in the conventional way, i.e.\ by computations without topology fixing; $\langle Q^2 \rangle / \hat{V}$: the topological susceptibility obtained via $\langle Q^2 \rangle / \hat{V}$).}
\end{table}


\section{Conclusions}

We successfully determined the pion mass and the static potential in the Schwinger model from computations at fixed topology using a method proposed and equations derived in \cite{Brower:2003yx}. The generalization of the method to QCD seems to be straightforward. There it might be used to circumvent problems associated with topology freezing expected at small values of the lattice spacing or when using e.g.\ overlap fermions.


\begin{acknowledgments}

We thank Carsten Urbach for providing program code for simulating the Schwinger model. We thank Wolfgang Bietenholz, Arthur Dromard, Gregorio Herdoiza, Karl Jansen and Carsten Urbach for discussions.

M.W.\ acknowledges support by the Emmy Noether Programme of the DFG (German Research Foundation), grant WA 3000/1-1.

This work was supported in part by the Helmholtz International Center for FAIR within the framework of the LOEWE program launched by the State of Hesse.

\end{acknowledgments}




\begin{thebibliography}{99}
  
\bibitem{Luscher:2011kk}
  M.~L\"uscher and S.~Schaefer,
  JHEP {\bf 1107}, 036 (2011)
  [arXiv:1105.4749 [hep-lat]].

\bibitem{Aoki:2008tq}
  S.~Aoki {\it et al.} [JLQCD Collaboration],
  Phys.\ Rev.\ D {\bf 78}, 014508 (2008)
  [arXiv:0803.3197 [hep-lat]].

\bibitem{Fukaya:2005cw} 
  H.~Fukaya {\it et al.},
  Phys.\ Rev.\ D {\bf 73}, 014503 (2006)
  [hep-lat/0510116].

\bibitem{Bietenholz:2005rd} 
  W.~Bietenholz {\it et al.},
  JHEP {\bf 0603}, 017 (2006)
  [hep-lat/0511016].

\bibitem{Bruckmann:2009cv} 
  F.~Bruckmann {\it et al.},
  Eur.\ Phys.\ J.\ A {\bf 43}, 303 (2010)
  [arXiv:0905.2849 [hep-lat]].

\bibitem{Brower:2003yx}
  R.~Brower {\it et al.},
  Phys.\ Lett.\ B {\bf 560}, 64 (2003)
  [hep-lat/0302005].

\bibitem{Aoki:2007ka}
  S.~Aoki {\it et al.},
  Phys.\ Rev.\ D {\bf 76}, 054508 (2007)
  [arXiv:0707.0396 [hep-lat]].

\bibitem{Bietenholz:2012sh} 
  W.~Bietenholz and I.~Hip,
  J.\ Phys.\ Conf.\ Ser.\ {\bf 378}, 012041 (2012)
  [arXiv:1201.6335 [hep-lat]].

\bibitem{Dromard:2013wja} 
  A.~Dromard and M.~Wagner,
  arXiv:1309.2483 [hep-lat].

\bibitem{Smilga:1996pi}
  A.~V.~Smilga,
  Phys.\ Rev.\ D {\bf 55} (1997) 443
  [hep-th/9607154].

\bibitem{Gattringer:1999gt}
  C.~Gattringer, I.~Hip and C.~B.~Lang,
  Phys.\ Lett.\ B {\bf 466} (1999) 287
  [hep-lat/9909025].

\bibitem{Luscher:1981zq} 
  M.~L\"uscher,
  Commun.\ Math.\ Phys.\ {\bf 85}, 39 (1982).

\bibitem{Gattringer:1997qc} 
  C.~R.~Gattringer, I.~Hip and C.~B.~Lang,
  Nucl.\ Phys.\ B {\bf 508}, 329 (1997)
  [hep-lat/9707011].

\bibitem{HMC_Urbach}
  \texttt{https://github.com/urbach/schwinger}.

\bibitem{Coleman:1978ae} 
  S.~R.~Coleman,
  Subnucl.\ Ser.\ {\bf 15}, 805 (1979).
  
\end{thebibliography}
\end{document}